\documentclass[runningheads]{llncs}
\usepackage{graphicx}
\usepackage{cite}
\usepackage{hyperref}
\usepackage{breakcites}
\usepackage{microtype}
\usepackage{subcaption}
\usepackage{romannum}
\usepackage{amsmath}
\usepackage{mathtools}
\usepackage{amssymb}
\usepackage{booktabs}
\usepackage[most]{tcolorbox}
\usepackage{etoolbox} 
\usepackage{wrapfig,lipsum,booktabs} 

\makeatletter
\patchcmd{\ps@headings}{\rlap{\thepage}}{}{}{}
\patchcmd{\ps@headings}{\llap{\thepage}}{}{}{}
\makeatother
\begin{document}
\title{Image Synthesis as a Pretext for Unsupervised Histopathological Diagnosis}

\author{Dejan Štepec\inst{1,2} \and Danijel Skočaj\inst{1}}

\institute{University of Ljubljana, Faculty of Computer and Information Science\\
Večna pot 113, 1000 Ljubljana, Slovenia
\and
XLAB d.o.o.\\
Pot za Brdom 100, 1000, Ljubljana, Slovenia\\
\email{dejan.stepec@xlab.si}}

\maketitle

\begin{abstract}

Anomaly detection in visual data refers to the problem of differentiating abnormal appearances from normal cases. Supervised approaches have been successfully applied to different domains, but require abundance of labeled data. Due to the nature of how anomalies occur and their underlying generating processes, it is hard to characterize and label them. Recent advances in deep generative based models have sparked interest towards applying such methods for unsupervised anomaly detection and have shown promising results in medical and industrial inspection domains. In this work we evaluate a crucial part of the unsupervised visual anomaly detection pipeline, that is needed for normal appearance modelling, as well as the ability to reconstruct closest looking normal and tumor samples. We adapt and evaluate different high-resolution state-of-the-art generative models from the face synthesis domain and demonstrate their superiority over currently used approaches on a challenging domain of digital pathology. Multifold improvement in image synthesis is demonstrated in terms of the quality and resolution of the generated images, validated also against the supervised model.

\end{abstract}

\section{Introduction}\raggedbottom

Anomaly detection represents an important process of determining instances that stand out from the rest of the data. Detecting such occurrences in different data modalities is widely applicable in different domains such as fraud detection, cyber-intrusion, industrial inspection and medical imaging~\cite{anomaly_survey}. Detecting anomalies in high-dimensional data (e.g. images) is a particularly challenging problem, that has recently seen a particular rise of interest, due to prevalence of deep-learning based methods, but their success has mostly relied on abundance of available labeled data.

Anomalies generally occur rarely, in different shapes and forms and are thus extremely hard or even impossible to label. Supervised deep-learning approaches have seen great success, especially evident in the domains with well known characterization of the anomalies and abundance of labeled data. Obtaining such detailed labels to learn supervised models is a costly and in many cases also an impossible process, due to unknown set of all the disease biomarkers or product defects. In an unsupervised setting, only normal samples are available (e.g. healthy, defect-free), without any labels. Deep generative methods have been recently applied to the problem of unsupervised anomaly detection (UAD), by utilizing the abundance of unlabeled data and demonstrating promising results in medical and industrial inspection domains~\cite{f-anogan,anomaly_brain, mvtec_ad}. Deep generative methods, in a form of autoencoders~\cite{autoencoder} or GANs~\cite{gan_org} are in a UAD setting used to capture normal appearance, in order to detect and segment deviations from that normal appearance, without the need for labeled data.

AnoGAN~\cite{anogan} represents the first method, where GANs are used for anomaly detection in medical domain. A rich generative model is constructed on healthy examples of optical coherence tomography (OCT) images of the retina and a methodology is presented for image mapping into the latent space. The induced latent vector is used to generate the closest example to the presented query image, in order to detect and segment the anomalies in an unsupervised fashion. AnoGAN~\cite{anogan} and the recently presented f-AnoGAN~\cite{f-anogan} improvement, utilize low resolution vanilla DCGAN~\cite{DCGAN} and Wasserstein GAN~\cite{WGAN} architectures, for normal appearance modelling, with significantly lower anomaly detection performance in comparison with autoencoder-based approaches~\cite{ganomaly, skip_ganomaly}. This does not coincide with superior image synthesis performance of the recent GAN-based methods. We argue, that adapting recent advancements in GAN-based unconditional image generation~\cite{pgan, styleGAN, styleGAN2}, currently utilized mostly for human face synthesis, should also greatly improve the performance of image synthesis in different medical imaging domains, as well GAN-based UAD methods.

In this work we focus on normal appearance modelling part of the UAD pipeline and evaluate the ability to synthesize realistically looking histology imagery. This presents a pretext for an important problem of metastases detection from digitized gigapixel histology imagery. This particular problem has been already addressed in a supervised setting~\cite{jama_dl}, by relying on the limited amount of expertly lesion-level labeled data, as well as in a weakly-supervised setting~\cite{weakly_nature}, where only image-level labels were used. Extremely large histology imagery and highly variable appearance of the anomalies (i.e. cancerous regions) represent a unique challenge for existing UAD approaches. We investigate the use of the recently presented high resolution generative models from the human face synthesis domain~\cite{pgan, styleGAN, styleGAN2}, for normal appearance modelling in a UAD pipeline, which could consequently improve the performance and stability of the current state-of-the-art approaches~\cite{f-anogan, skip_ganomaly}. 

We demonstrate this with significant improvements in the quality and increased resolution of the generated imagery in comparison with currently used approaches~\cite{DCGAN, WGAN}, which also represents a novel application of generative models to the digital pathology domain. We also investigate the effectiveness of current latent space mapping approaches, specifically their ability of closest looking normal histology sample reconstruction. We validate the quality of the synthesized imagery against the supervised model and demonstrate the importance of synthesizing high resolution histology imagery, resulting in an increased amount of contextual information present, crucial for distinguishing tumor samples from the normal ones.

\section{Methodology}
\subsubsection{Unsupervised Visual Anomaly Detection} 
\begin{figure*}[ht!]
    \centering
    \begin{subfigure}{0.45\textwidth}
      \includegraphics[width=\linewidth]{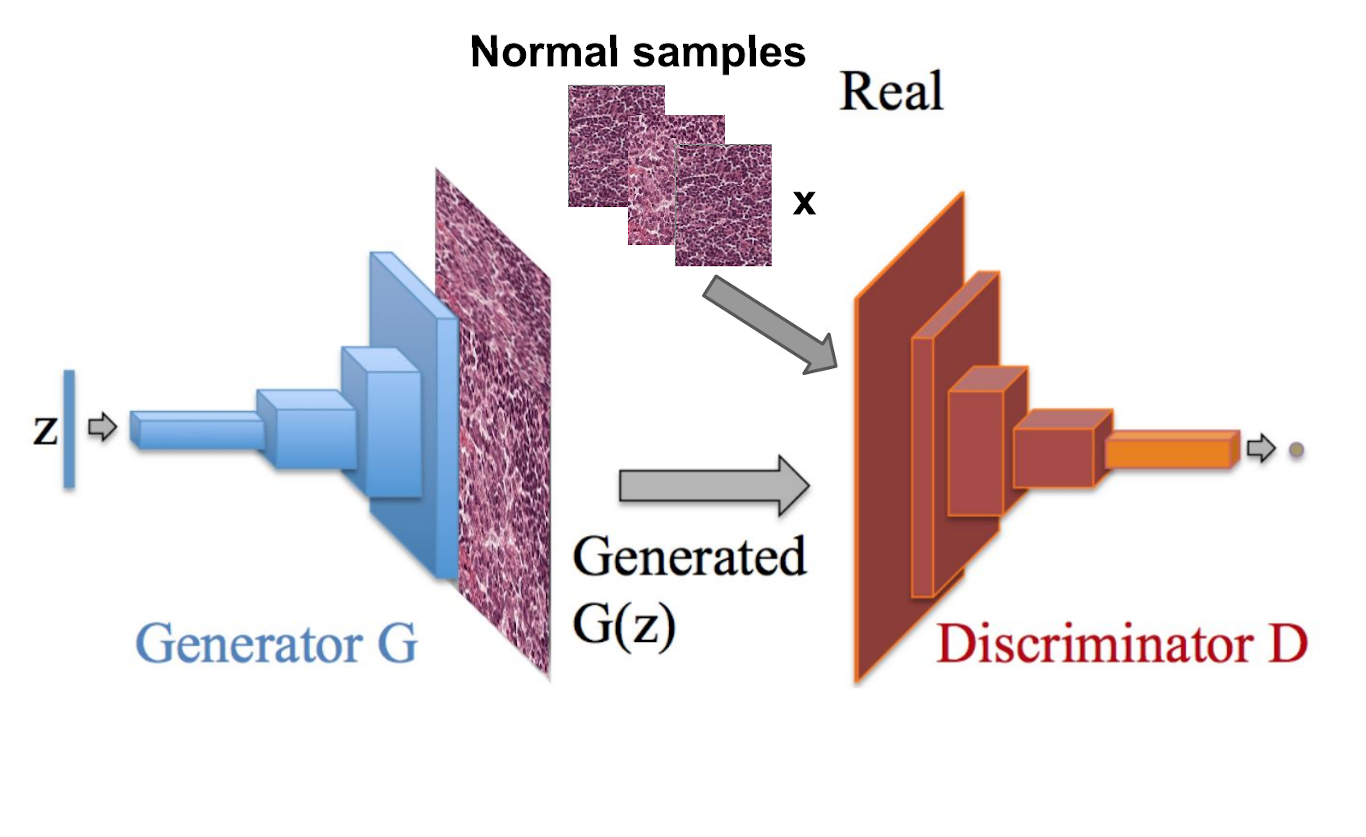}
      \caption{Normal appearance modelling}
      \label{fig:anogan_gan}
    \end{subfigure}\hfil
        \begin{subfigure}{0.44\textwidth}
      \includegraphics[width=\linewidth]{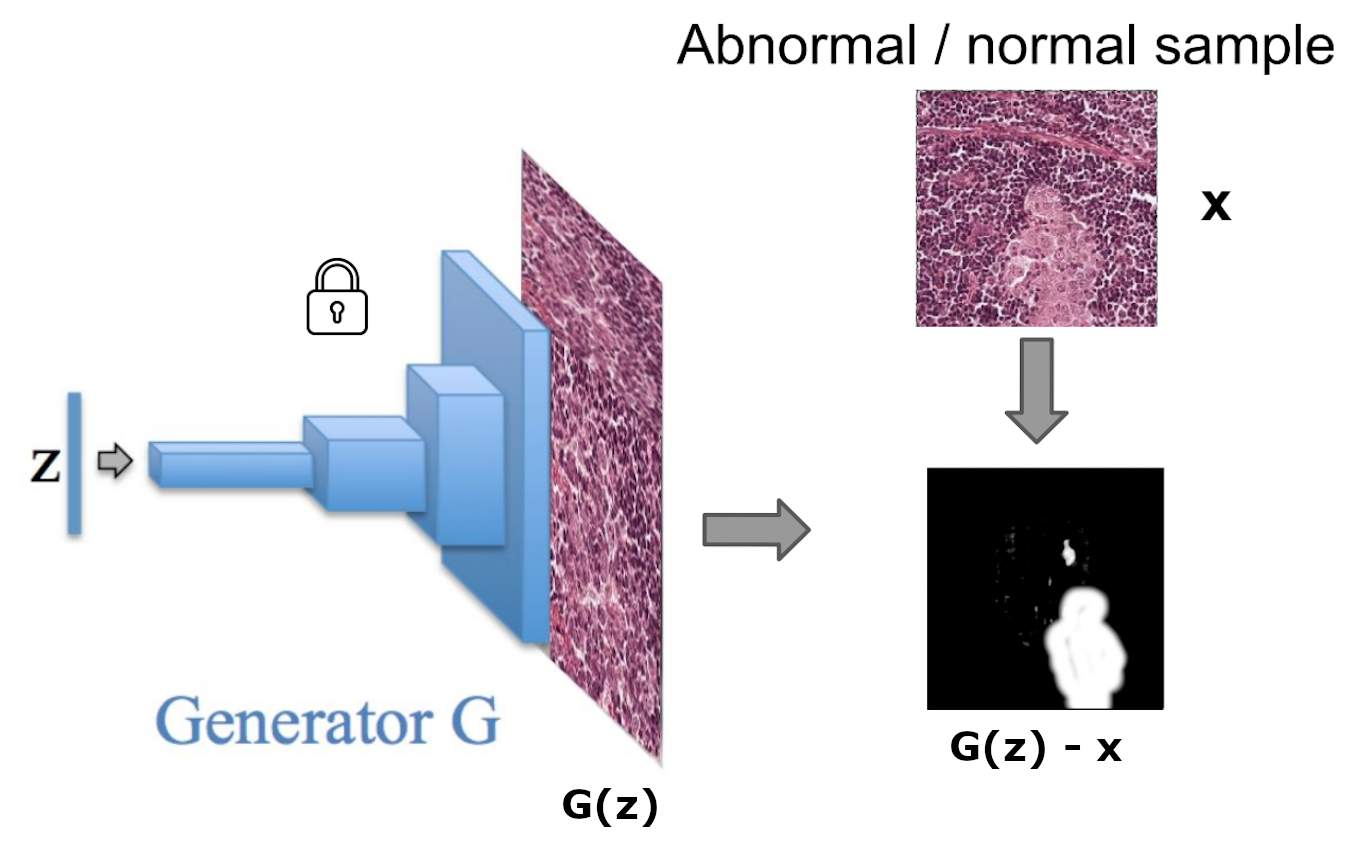}
      \caption{Anomaly detection}
      \label{fig:anogan_anomaly}
    \end{subfigure}\hfil
    \caption{GAN based visual anomaly detection pipeline, consisting out of a) normal appearance modelling and  b) the search for optimal latent representation, that will generate the closest normal appearance sample, used for anomaly detection.}
    \label{fig:anogan}
\end{figure*}

The capability to learn the distribution of the normal appearance represents one of the most important parts in a visual anomaly detection pipeline, presented in Figure~\ref{fig:anogan}. This is achieved by learning deep generative models on normal samples only, as presented in Figure~\ref{fig:anogan_gan}. The result of this process is the capability to generate realistically looking artificial normal samples, which cannot be distinguished from the real ones. To detect an anomaly, a query image is presented and the closest possible normal sample appearance is generated, which is used to threshold the difference, in order to detect and segment the anomalous region, as presented in Figure~\ref{fig:anogan_anomaly}. This is possible due to learned manifold of normal appearance and its inability to reconstruct anomalous samples.

Different approaches have been proposed for normal appearance modelling, as well as anomaly detection. Learning the normal visual appearance is based on autoencoders~\cite{anomaly_brain}, GANs~\cite{anogan, f-anogan}, or combined hybrid models~\cite{ganomaly, skip_ganomaly}. Most of the approaches learn the space of the normal sample distribution $Z$, from which latent vectors $z \in Z$ are sampled from, that generate the closest normal appearance, to the presented query image. Different solutions have been proposed for latent vector optimization, that are usually independent from the used normal appearance modelling method (i.e. autoencoders, GANs).

Current state-of-the-art GAN-based visual anomaly detection methods~\cite{anogan, f-anogan} are based on vanilla GAN implementations~\cite{DCGAN, WGAN}, with limited resolution generated normal samples of low fidelity, as well as with stability problems. Autoencoder based anomaly detection methods have shown improved results over pure GAN based implementations~\cite{ganomaly, skip_ganomaly}, but are similarly limited to a low resolution of $64^2$. In comparison, we evaluate the feasibility of generating high fidelity histology imagery up to the resolution of $512^2$, with the recently presented GAN architectures~\cite{pgan, styleGAN, styleGAN2} from the face generation domain, not yet utilized in anomaly detection pipelines, as well as in the digital pathology domain. Note, that we limit the resolution to a maximum of $512^2$, due to hardware resource and time constraints. We also investigate the effectiveness of recently presented latent space mapping approaches and the feasibility to be applicable for UAD in the digital pathology domain.

\subsubsection{Deep Generative Adversarial Models.} Original GAN implementation~\cite{gan_org} was based on standard neural networks, which generated images suffered from being noisy and the training process was notoriously unstable. This was improved by implementing the GAN idea using the CNNs - DCGAN~\cite{DCGAN}, by identifying a family of architectures, that result in a stable training process of higher resolution deep generative models. The method was later on also adapted in the AnoGAN anomaly detection framework~\cite{anogan}. Stability problems of GAN methods were first improved by proposing different distance measures for the cost function (e.g. Wasserstein GAN~\cite{WGAN}), adapted also by the f-AnoGAN anomaly detection method~\cite{f-anogan}. The main limitation of those early GAN methods is also the low resolution (up to $64^2$) and the limited variability of the generated images.

Recently, the ideas of progressively growing GANs~\cite{pgan} and style-based generators~\cite{styleGAN, styleGAN2} were presented, allowing a stable training of models for resolutions up to $1024^2$, with increased variation and quality of the generated images. In progressively growing GANs~\cite{pgan}, layers are added to the generator and discriminator progressively, by linearly fading them in and thus enabling fast and stable training. StyleGAN~\cite{styleGAN} proposes an alternative generator architecture, based on style transfer literature~\cite{style_transfer}, exposing novel ways to control synthesis process and reducing the entanglement of the latent space. StyleGAN2~\cite{styleGAN2} addresses some of the main characteristic artifacts resulting from the progressive growing in StyleGAN~\cite{styleGAN}, further boosting generative performance.

In comparison with autoencoders, GANs do not automatically yield the inverse mapping from the image to latent space, which is needed for closest-looking normal sample reconstruction and consequently anomaly detection. In AnoGAN~\cite{anogan} an iterative optimization approach was proposed to optimize the latent vector $z$ via backpropagation, using the residual and discrimination loss. Residual loss is represented with pixel-wise Mean Square Error (MSE) loss, while discrimination loss is guided by the GAN discriminator, by computing feature matching loss between the real and synthesized imagery. In f-AnoGAN method~\cite{f-anogan}, an autoencoder replaces the iterative optimization procedure, using the trainable encoder and the pre-trained generator, as the decoder. For StyleGAN2~\cite{styleGAN2}, authors proposed an iterative inverting procedure, which specifically optimizes an intermediate latent space and noise maps, based on the Learned Perceptual Image Patch Similarity (LPIPS)~\cite{lpips}.

\section{Experiments and Results}
\label{sec:results}

\subsubsection{Histology Imagery Dataset.} 
\begin{figure*}[htbp]
    \centering
    \begin{subfigure}{0.33\textwidth}
      \includegraphics[width=\linewidth]{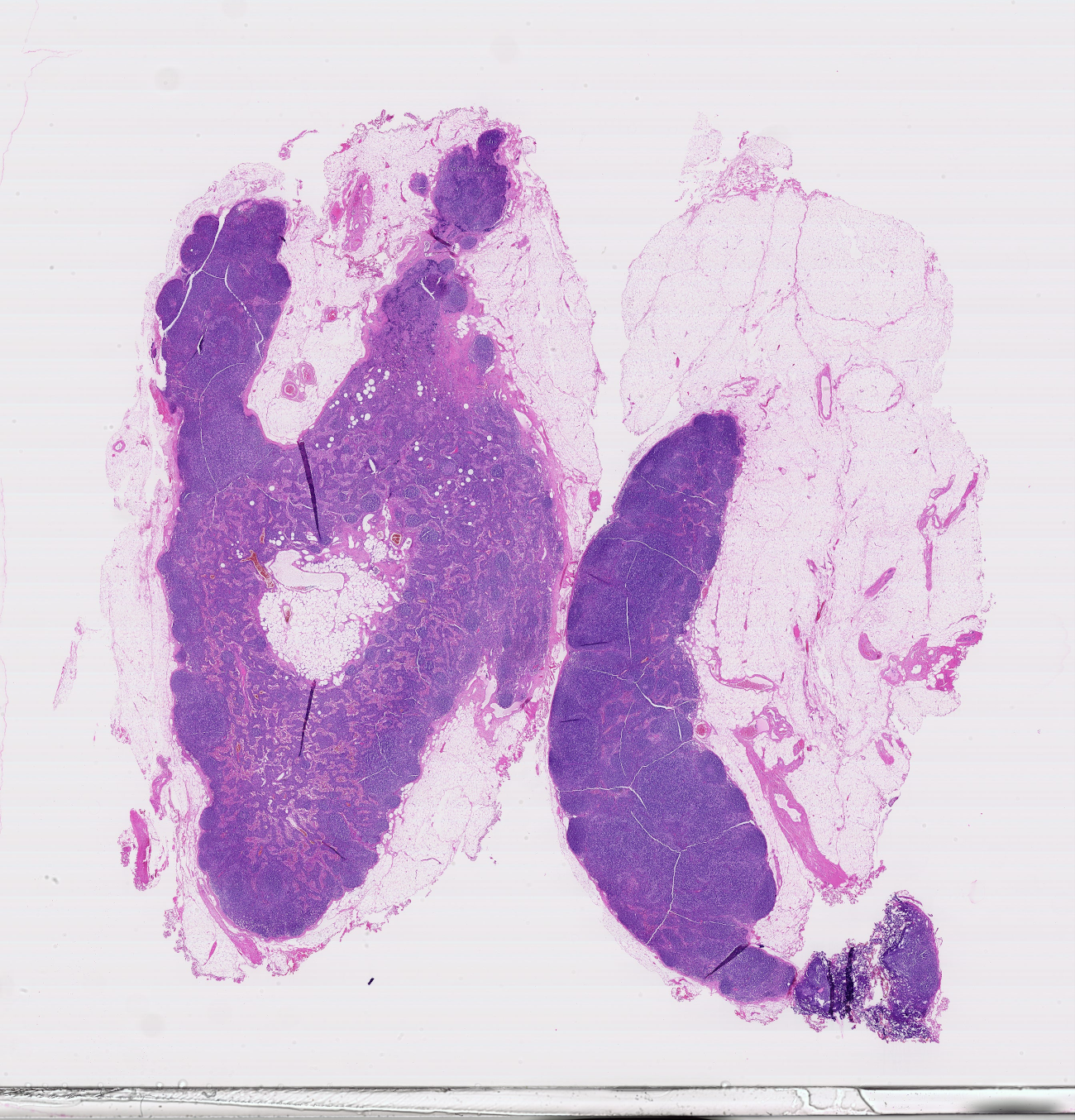}
      \caption{Original WSI}
      \label{fig:original_wsi}
    \end{subfigure}\hfil
        \begin{subfigure}{0.33\textwidth}
      \includegraphics[width=\linewidth]{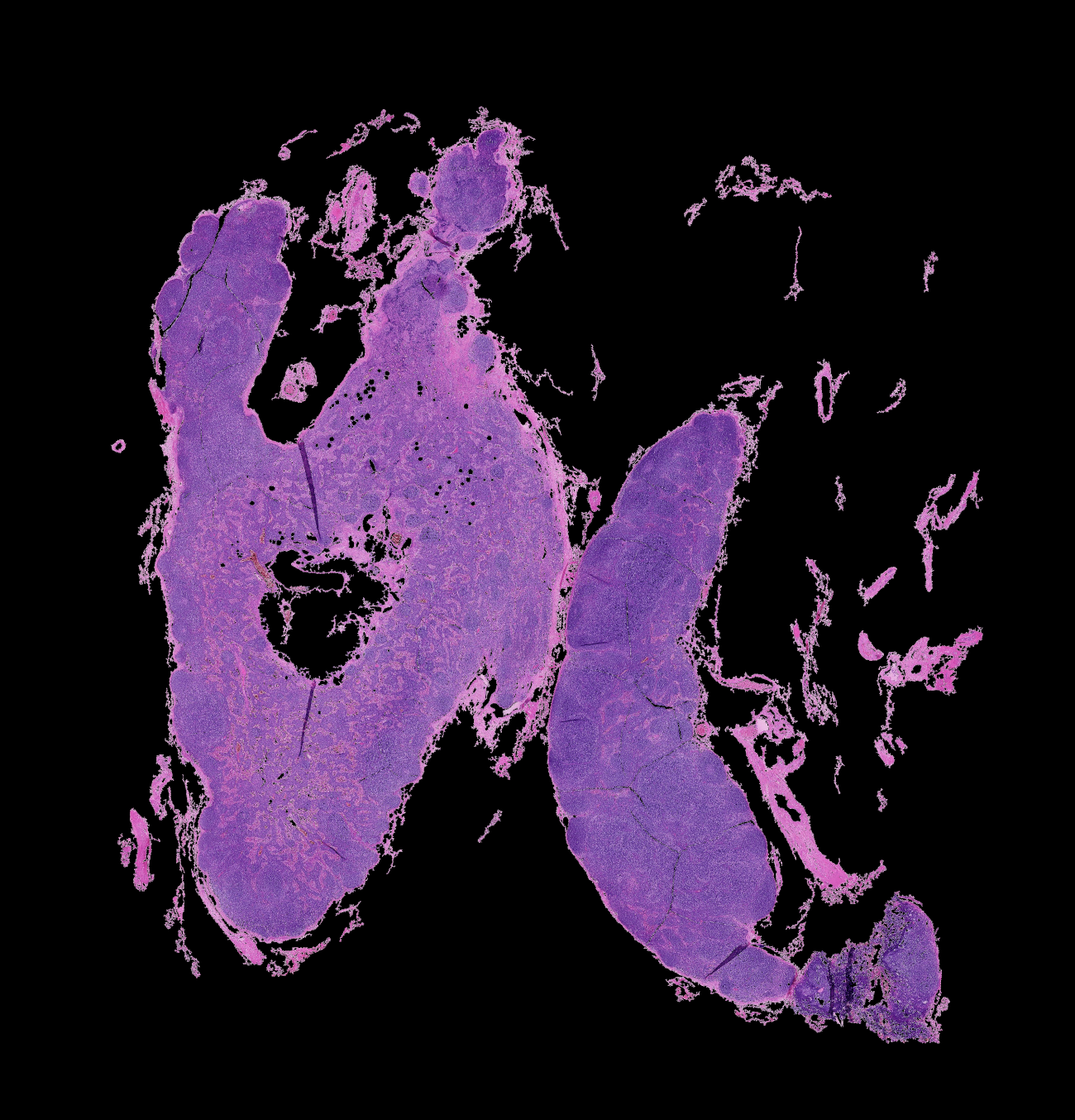}
      \caption{Filtered WSI}
      \label{fig:filtered_wsi}
    \end{subfigure}\hfil
    \begin{subfigure}{0.33\textwidth}
      \includegraphics[width=\linewidth]{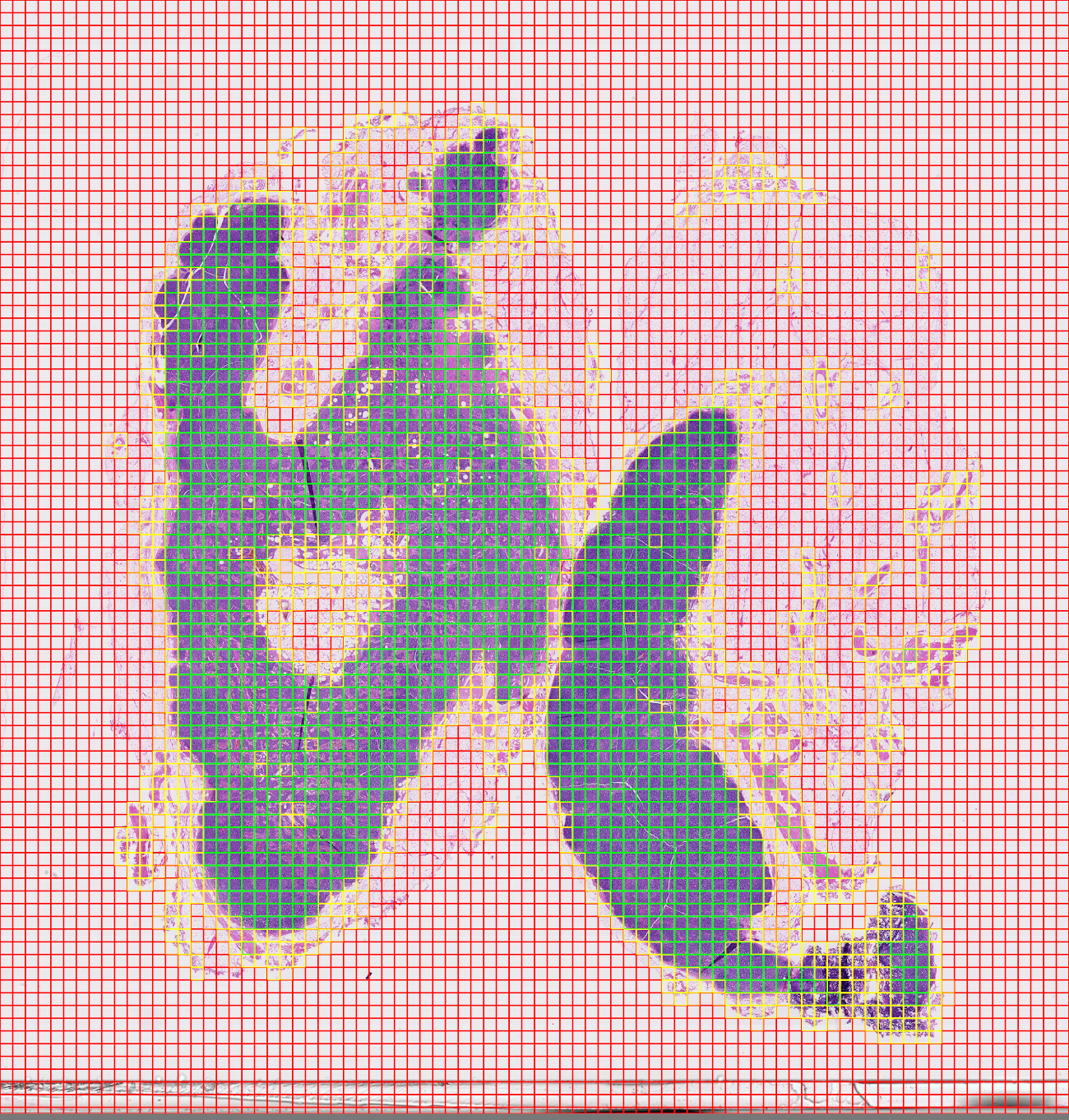}
      \caption{Patches from WSI ($1024^2$)}
      \label{fig:patches_wsi}
    \end{subfigure}\hfil
    \caption{Preprocessing of the original WSI presented in a) consists of b) filtering tissue sections and c) extracting patches, based on tissue percentage (green $\geq$ 90\%, red $\leq$ 10\% and yellow in-between). Best viewed in digital version with zoom.}
    \label{fig:preprocessing_wsi}
\end{figure*}

We address aforementioned problems of anomaly detection pipeline on a challenging domain of digital pathology, where whole-slide histology images (WSI) are used for diagnostic assessment of the spread of the cancer. This particular problem was already addressed in a supervised setting~\cite{jama_dl}, as a competition\footnote{https://camelyon16.grand-challenge.org/}, with provided clinical histology imagery and ground truth data. A training dataset with (n=110) and without (n=160) cancerous regions is provided, as well as a test set of 129 images (49 with and 80 without anomalies). Raw histology imagery, presented in Figure~\ref{fig:original_wsi}, is first preprocessed, in order to extract the tissue region (Figure~\ref{fig:filtered_wsi}). We used the approach from IBM\footnote{https://github.com/CODAIT/deep-histopath}, which utilizes a combination of morphological and color space filtering operations. Patches of different sizes ($64^2$ - $512^2$) are then extracted from the filtered image (Figure~\ref{fig:patches_wsi}) and labelled according to the amount of tissue in the extracted patch.

\subsubsection{Image Synthesis.} We first evaluate the performance of different GAN based generative approaches on a challenging histology imagery using the Fr\'echet Inception Distance (FID)~\cite{fid_score}, by following evaluation procedure from StyleGAN2~\cite[Table 1]{styleGAN2}. The FID score represents a similarity between a set of real and generated images, based on the statistics extracted from the pretrained Inception classifier~\cite{inception3}. We train different GANs using 1000 randomly extracted patches, from each of the 160 normal WSIs, with tissue coverage over 90\% (Figure~\ref{fig:patches_wsi}). This presents an input to the baseline DCGAN~\cite{DCGAN} model, used in the AnoGAN~\cite{anogan} anomaly detection framework, Wasserstein GAN (WGAN), used in f-AnoGAN~\cite{f-anogan}, as well as to recently presented GAN architectures, based on progressive growing (PGAN~\cite{pgan}) and style transfer (StyleGAN~\cite{styleGAN}, StyleGAN2~\cite{styleGAN2}). We evaluate not only the feasibility to generate pathology imagery, but to generate it up to a resolution of $512^2$ and present the results in Table~\ref{tbl:fid_results}.

\setlength{\tabcolsep}{7pt}
\begin{table*}[ht!]
    \centering
    \caption{FID scores for different methods and different input image sizes DCGAN and WGAN are limited to a maximum resolution of $64^2$ and only best performing model is evaluated at $512^2$.}
    \begin{tabular}{c|c|c|c|c|c}
    \toprule
Image size & DCGAN & WGAN & PGAN & StyleGAN & StyleGAN2 \\
\midrule
        $64^2$ &    88.52 & 12.65 &  18.89 &     7.15 &    \textbf{6.64} \\ \hline
        $256^2$ &    -&     -&       17.82 &     5.57 &    \textbf{5.24} \\ \hline
        $512^2$ &    -&     -&       -&          -&        \textbf{2.93} \\
\bottomrule
\end{tabular}
\label{tbl:fid_results}
\end{table*}

\begin{figure}
\centering
\begin{tcbraster}[raster columns=8, raster equal height, 
raster column skip=0pt, raster row skip=0pt, raster every box/.style={blank}]
\tcbincludegraphics{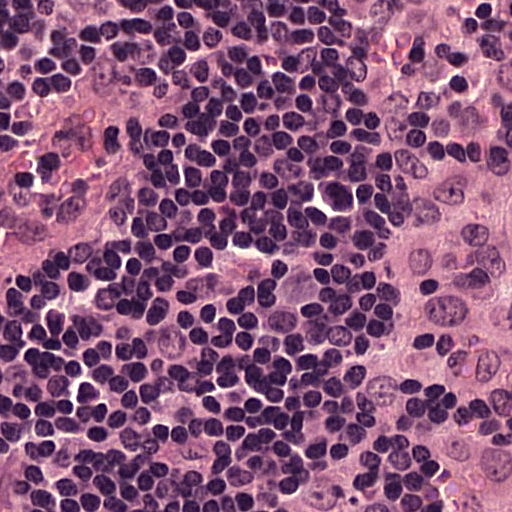}
\tcbincludegraphics{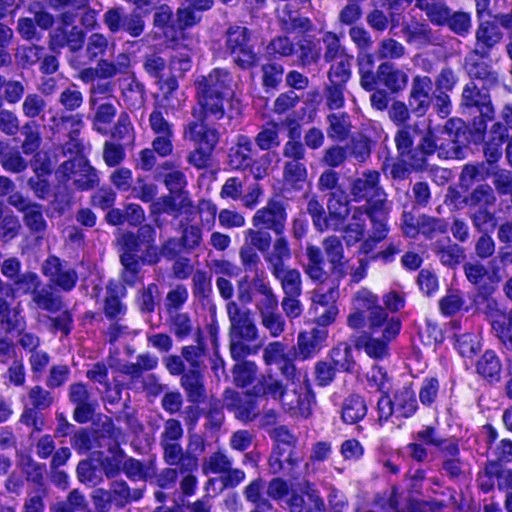}
\tcbincludegraphics{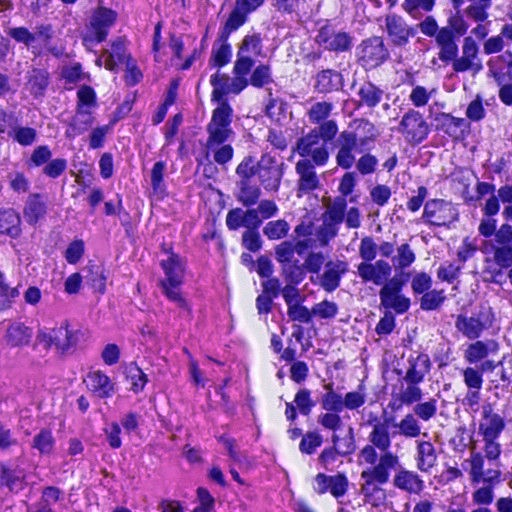}
\tcbincludegraphics{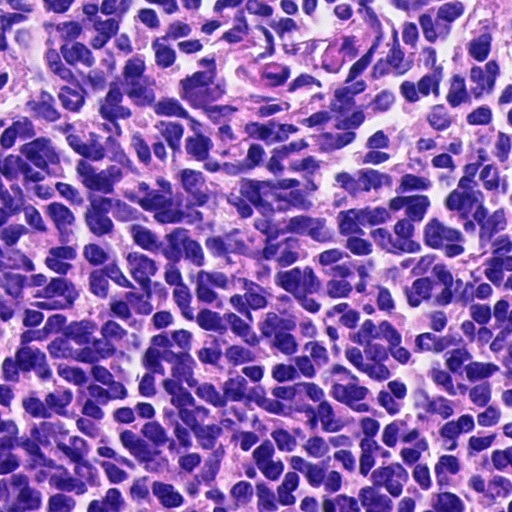}
\tcbincludegraphics{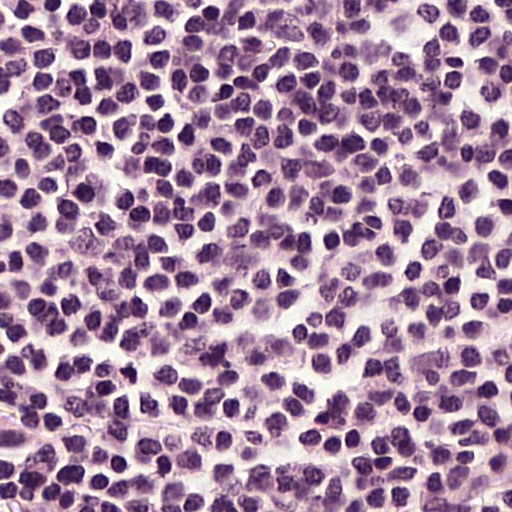}
\tcbincludegraphics{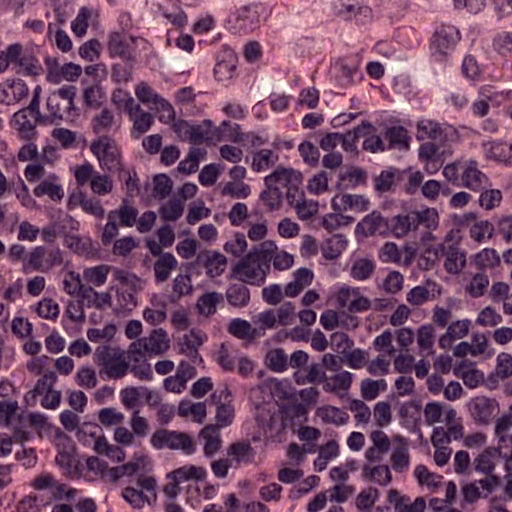}
\tcbincludegraphics{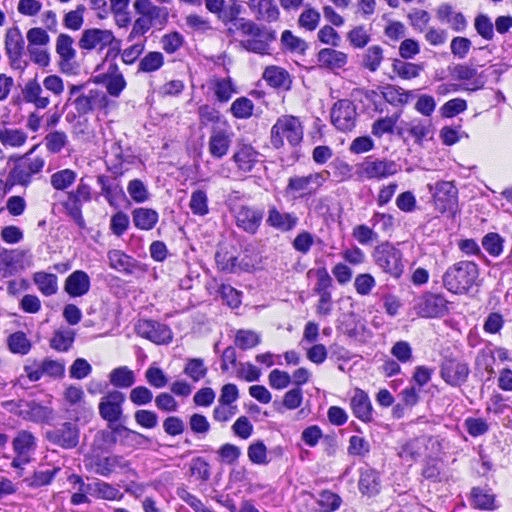}
\tcbincludegraphics{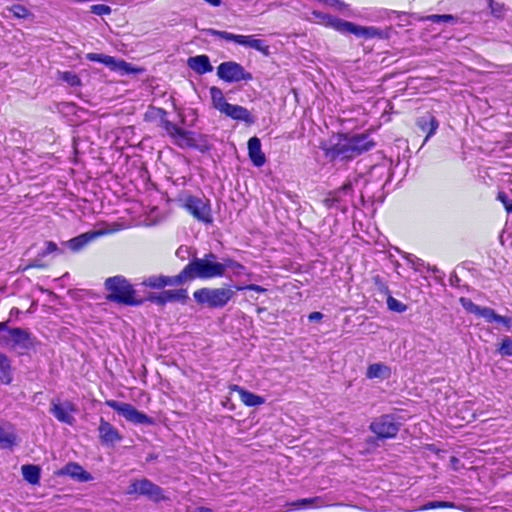}
\tcbincludegraphics{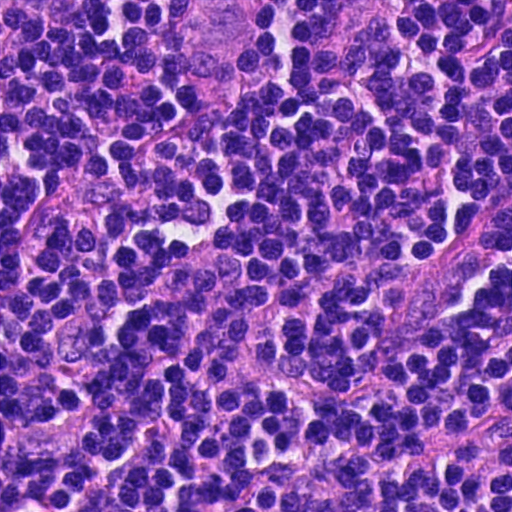}
\tcbincludegraphics{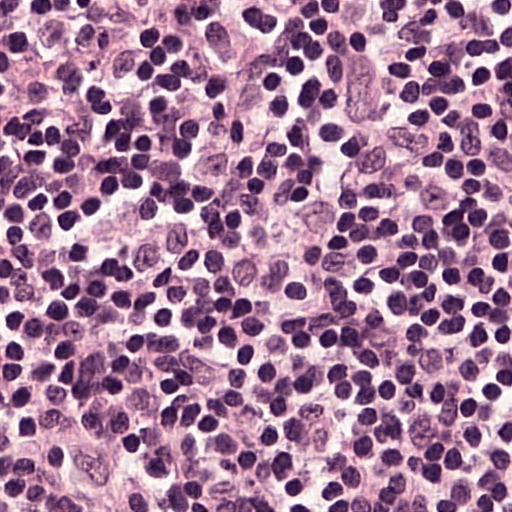}
\tcbincludegraphics{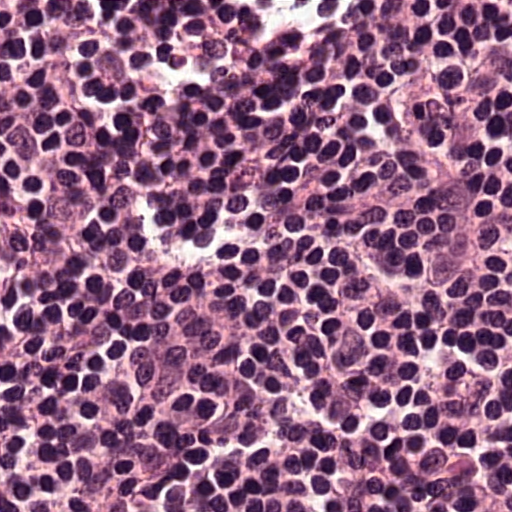}
\tcbincludegraphics{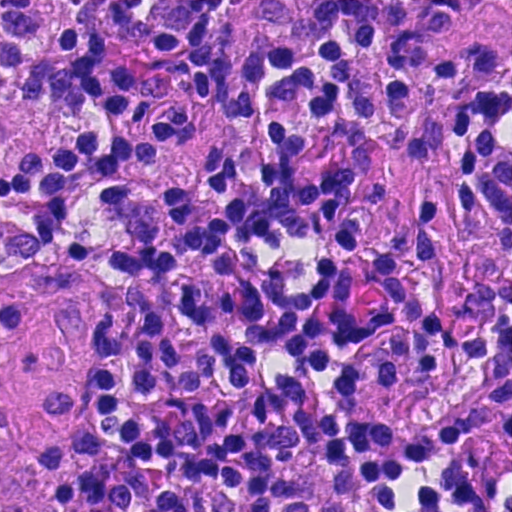}
\tcbincludegraphics{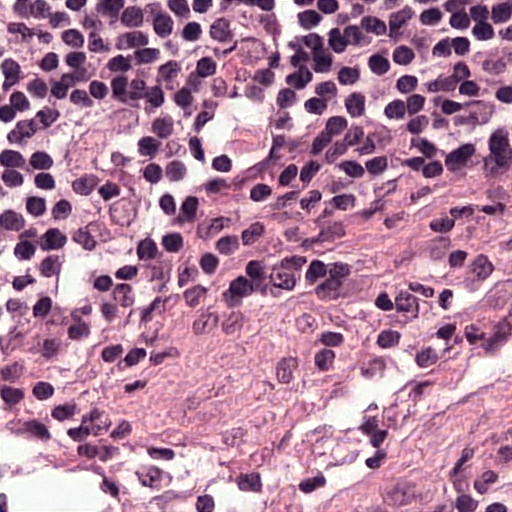}
\tcbincludegraphics{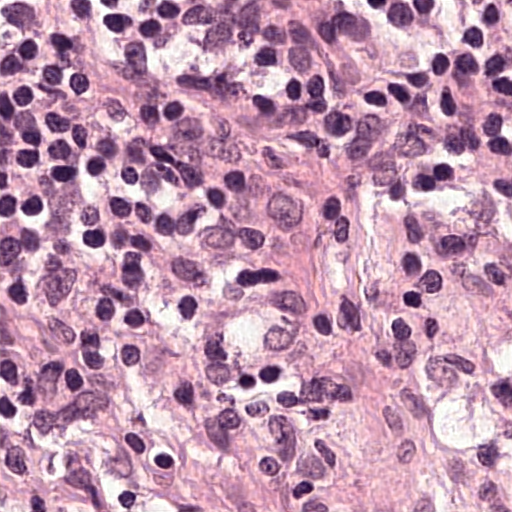}
\tcbincludegraphics{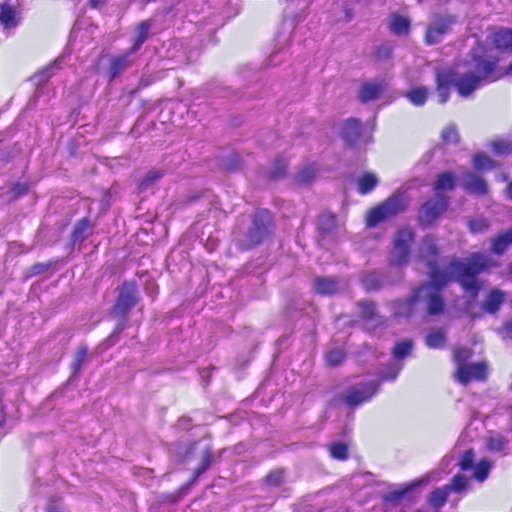}
\tcbincludegraphics{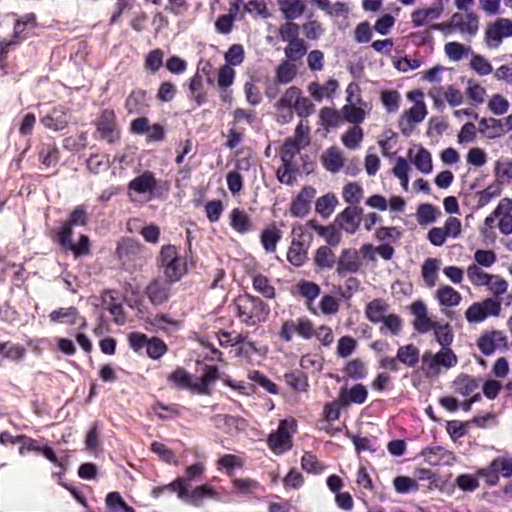}
\tcbincludegraphics{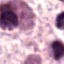}
\tcbincludegraphics{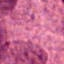}
\tcbincludegraphics{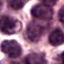}
\tcbincludegraphics{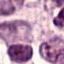}
\tcbincludegraphics{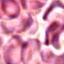}
\tcbincludegraphics{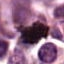}
\tcbincludegraphics{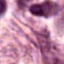}
\tcbincludegraphics{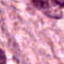}
\tcbincludegraphics{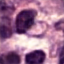}
\tcbincludegraphics{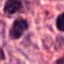}
\tcbincludegraphics{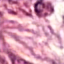}
\tcbincludegraphics{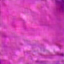}
\tcbincludegraphics{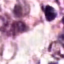}
\tcbincludegraphics{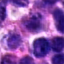}
\tcbincludegraphics{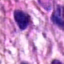}
\tcbincludegraphics{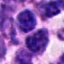}
\end{tcbraster}
\caption{Examples of real histology imagery at image size of $512^2$ (top row), generated images by the best performing StyleGAN2 model at $512^2$ (second row), real histology imagery at $64^2$ (third row) and images generated by the WGAN model at $64^2$ (last row). Best viewed in digital version with zoom.}
\label{fig:visual_results}
\end{figure}

We can see that generative performance of the recently presented methods significantly outperforms vanilla DCGAN model and should be considered in all the proposed GAN based visual anomaly detection pipelines. They are also capable of generating images of much bigger size and higher resolution, which is particularly important for anomaly detection, due to the increased amount of visual context, available for determining the presence or absence of the anomalies. For WGAN, we used the implementation from f-AnoGAN~\cite{f-anogan}, based on residual neural networks, which also produces high quality, high fidelity imagery up to image size of $64^2$. There is no significant difference in terms of the FID score between different style transfer based approaches, but StyleGAN2 represents an incremental improvement over StyleGAN and also offers an additional benefit of generator reversibility, especially interesting for anomaly detection (Figure~\ref{fig:anogan_anomaly}). Note that in StyleGAN2 a smaller network configuration $E$ was used for an increased image throughput, at small performance expense~\cite[Table 1]{styleGAN2}. Visual comparison of generated results is presented in Figure~\ref{fig:visual_results}, demonstrating high variability and quality of the generated images and also the applicability of such methods for digital pathology.

\subsubsection{Classification of Real and Synthetic Patches.} To additionally asses the quality of the generated images in comparison with the real ones, we evaluated the performance of a discriminative classifier applied on both types of data. We trained supervised DenseNet121~\cite{densenet} model on extracted normal and tumor histology imagery patches and compared its performance to distinguish the two classes on real and synthesized imagery (Table~\ref{tbl:supervised_results}). We extracted 100.000 normal and tumor patches (with provided annotations) of size $64^2$ and $512^2$ to train DenseNet121 model and evaluated the performance on a test set of 10.000 (real) normal and tumor patches. Besides on normal histology imagery, we also trained WGAN~\cite{WGAN} and StyleGAN2~\cite{styleGAN2} on tumor patches, in order to be able to generate both classes. We then synthesized a test set of 10.000 normal and tumor patches and evaluated the capability of the DenseNet121 model (trained on real imagery) to distinguish anomalous samples in synthesized imagery.

\begin{table*}[ht!]
    \centering
    \caption{Classification accuracy (CA) for normal vs. tumor patch based classification on real and synthesized imagery (WGAN for $64^2$ and StyleGAN2 for $512^2$).}
    \label{tbl:supervised_results}
    \begin{tabular}{c|c|c}
        \toprule  
        Image size & Real (CA) & Synthesized (CA)\\  \midrule
        $64^2$ & 88.23\% & 87.05\% \\  \midrule
        $512^2$ & 98.89\% & 98.34\% \\  
        \bottomrule
    \end{tabular}
\end{table*}

Results in Table~\ref{tbl:supervised_results} demonstrate that supervised model (trained on real imagery) successfully recognizes synthesized imagery, with no drop in performance, compared to real imagery. Supervised model can be seen as a virtual pathologist, confirming to some extent the correctness of image synthesis, on a much larger scale. We can also see significant drop in performance on $64^2$ patches, which additionally confirms, that larger patches hold more contextual information, useful for classification.

\subsubsection{Latent Space Mapping.} 
\begin{figure*}[htbp]
    \centering
    \begin{subfigure}{0.48\textwidth}
      \begin{tcbraster}[raster columns=4, raster equal height, 
        raster column skip=0pt, raster row skip=0pt, raster every box/.style={blank}]
        \tcbincludegraphics{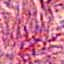}
        \tcbincludegraphics{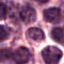}
        \tcbincludegraphics{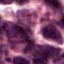}
        \tcbincludegraphics{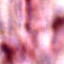}
        \tcbincludegraphics{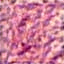}
        \tcbincludegraphics{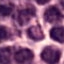}
        \tcbincludegraphics{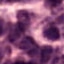}
        \tcbincludegraphics{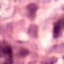}
        
        \tcbincludegraphics{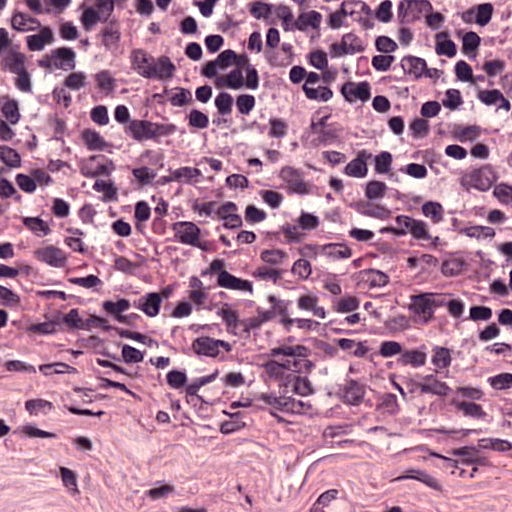}
        \tcbincludegraphics{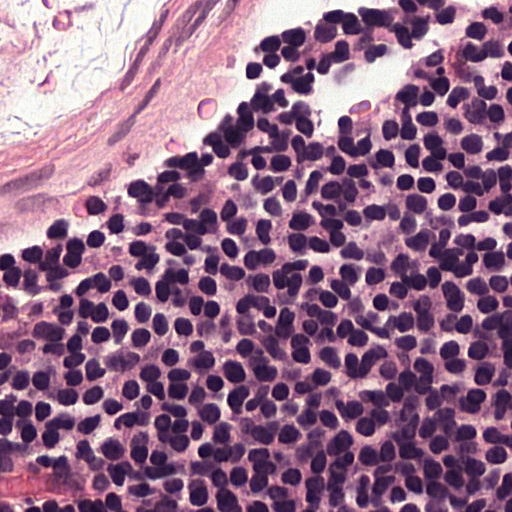}
        \tcbincludegraphics{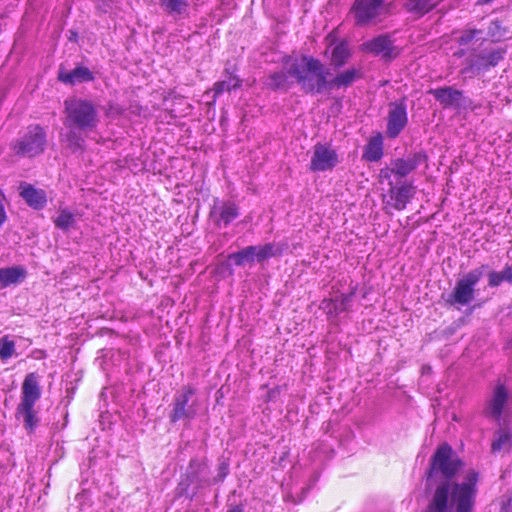}
        \tcbincludegraphics{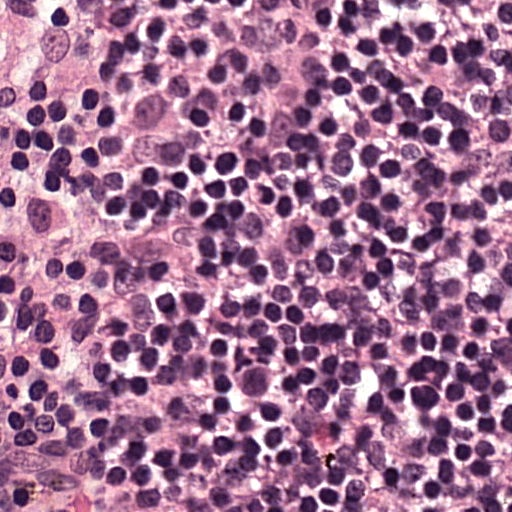}
        \tcbincludegraphics{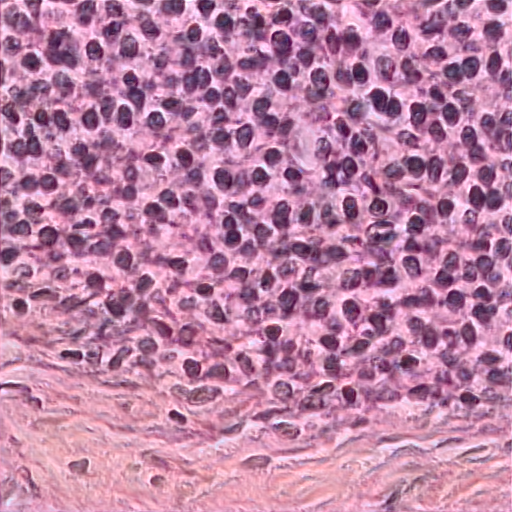}
        \tcbincludegraphics{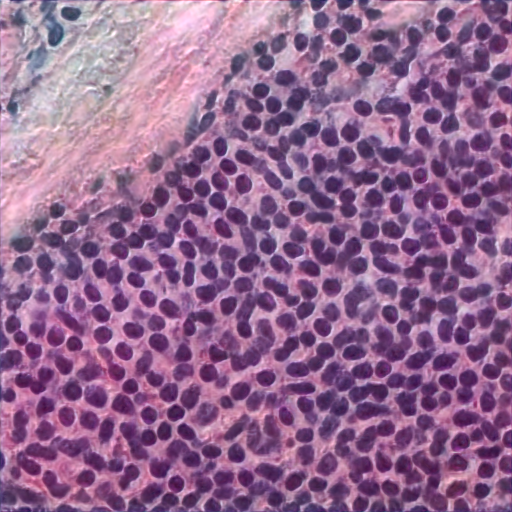}
        \tcbincludegraphics{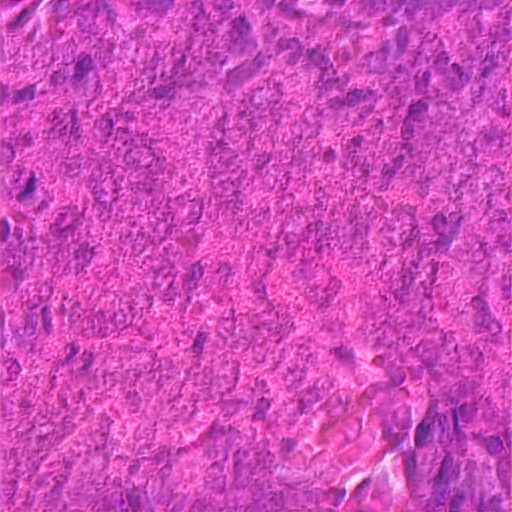}
        \tcbincludegraphics{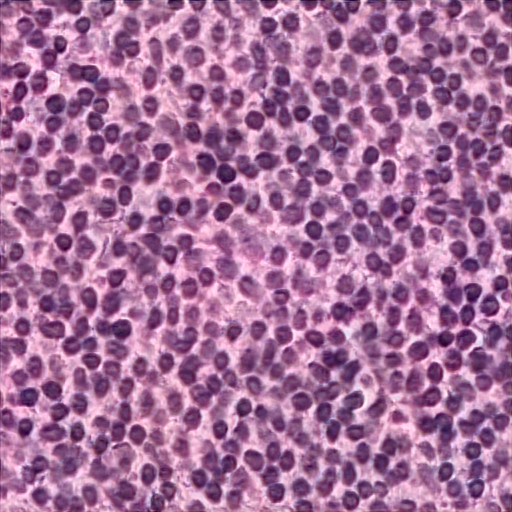}
    \end{tcbraster}
      \caption{normal}
      \label{fig:normal_an}
    \end{subfigure}\hfil
        \begin{subfigure}{0.48\textwidth}
      \begin{tcbraster}[raster columns=4, raster equal height, 
        raster column skip=0pt, raster row skip=0pt, raster every box/.style={blank}]
        \tcbincludegraphics{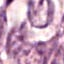}
        \tcbincludegraphics{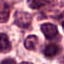}
        \tcbincludegraphics{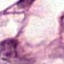}
        \tcbincludegraphics{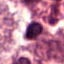}
        \tcbincludegraphics{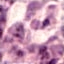}
        \tcbincludegraphics{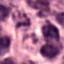}
        \tcbincludegraphics{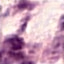}
        \tcbincludegraphics{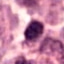}
        
        \tcbincludegraphics{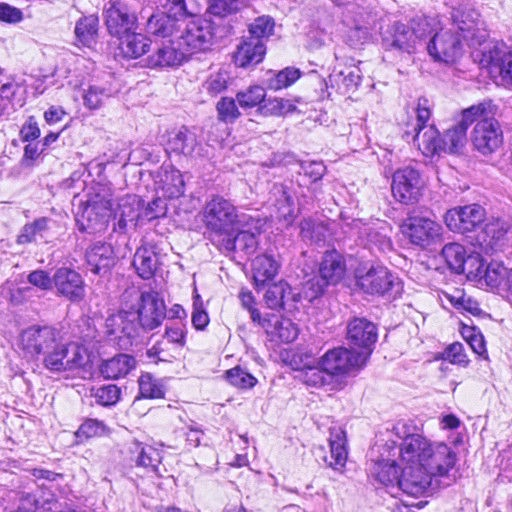}
        \tcbincludegraphics{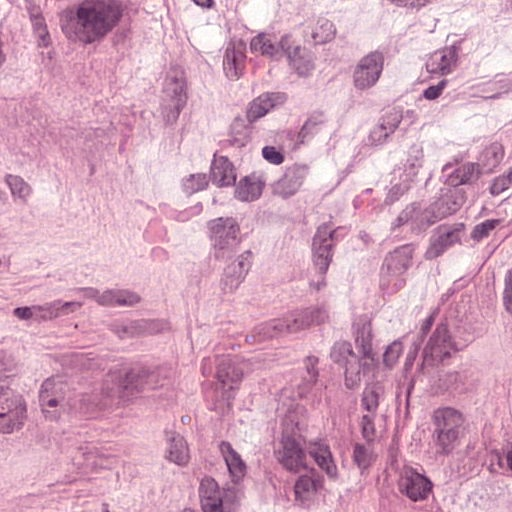}
        \tcbincludegraphics{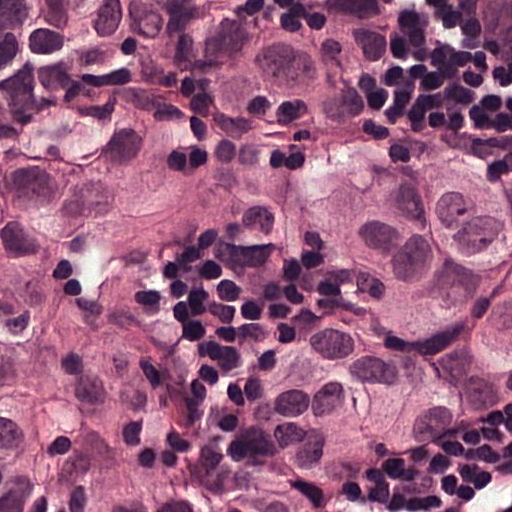}
        \tcbincludegraphics{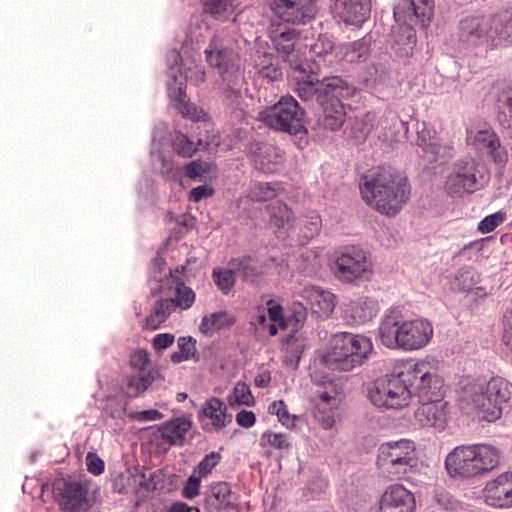}
        \tcbincludegraphics{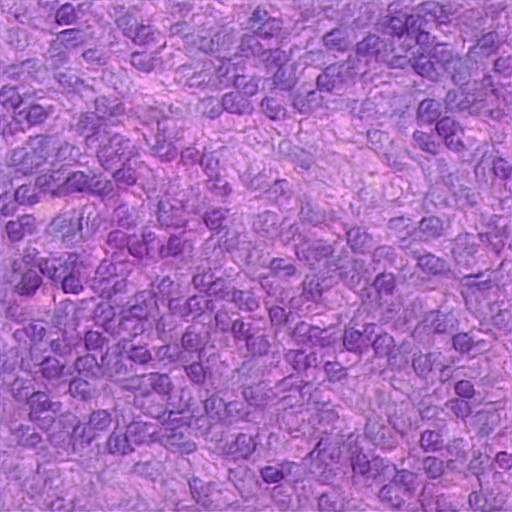}
        \tcbincludegraphics{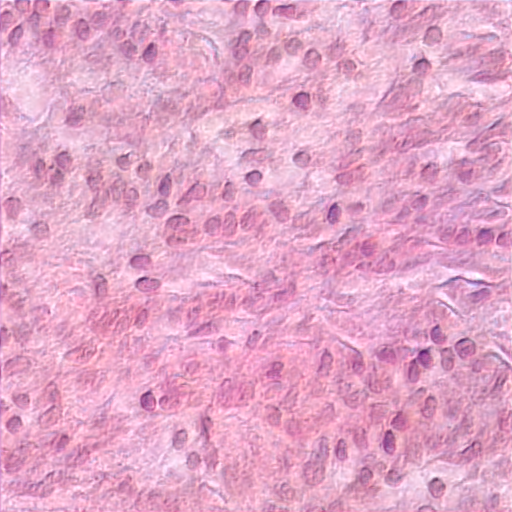}
        \tcbincludegraphics{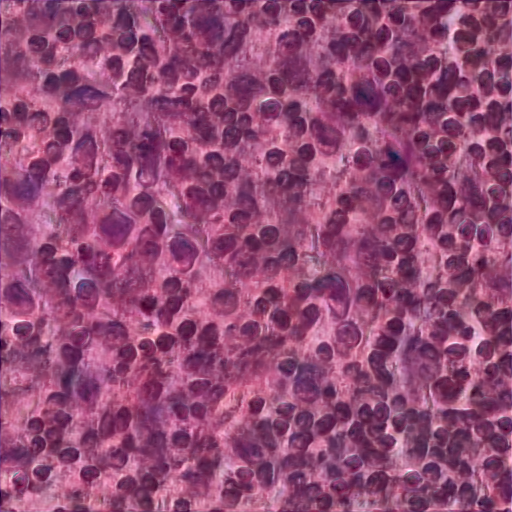}
        \tcbincludegraphics{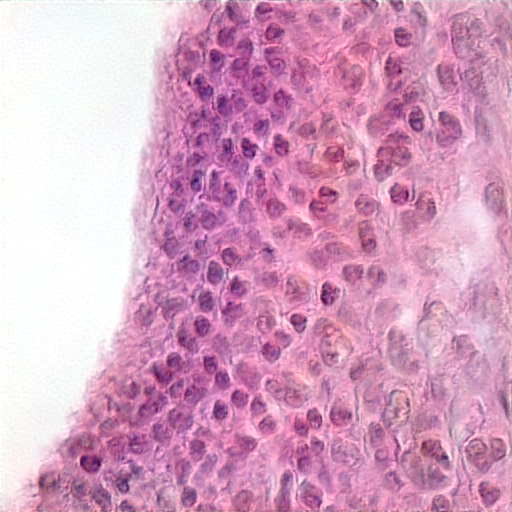}
    \end{tcbraster}
      \caption{tumor}
      \label{fig:backprojection}
    \end{subfigure}\hfil
    \caption{Projecting real a) normal and b) tumor imagery to the latent space of WGAN at $64^2$ (first row) and StyleGAN2 at $512^2$ (third row) and resulting closest matches in the latent space for WGAN (second row) and StyleGAN2 (last row). Best viewed in digital version with zoom.}
    \label{fig:preprocessing_wsi}
\end{figure*}

We qualitatively evaluate the capability of latent space projection using the encoder based approach ($izi_f$) presented in f-AnoGAN~\cite{f-anogan} for WGAN~\cite{WGAN} based generator and LPIPS-distance~\cite{lpips} based approach for StyleGAN2~\cite{styleGAN2} based generator, proposed and specifically designed for StyleGAN2 already in the original work~\cite{styleGAN2}. Figure~\ref{fig:backprojection} presents the results for both methods, image sizes and histopathological classes (i.e. normal and tumor). The generators are trained on normal samples only and should reconstruct only normal samples, while tumor samples should be poorly reconstructed, thus enabling detection of anomalies. We can see that the encoder based approach, proposed in f-AnoGAN~\cite{f-anogan} is able to find very similar looking artificial samples in the latent space. The problem is, that it also reconstructs tumor samples, with similar performance. We argue, that this is due to small patch size ($64^2$), which is in more than 10\% ambiguous also for the supervised classifier (Table~\ref{tbl:supervised_results}) - such small tumor patch might in fact not contain abnormality, or is represented with insufficient biomarkers. Such samples, even in small percentage~\cite{robust_ad2}, cause the generator to learn how to reconstruct anomalous samples.

StyleGAN2~\cite{styleGAN2} did not yield good reconstructions, capturing only major properties of the query image. This is beneficial for tumor samples, where we noticed consistent failure to capture even the main properties of the query image, with notable exception of the staining color. This can probably be attributed to the larger image sizes and more contextual information present to distinguish anomalous samples, not seen during the generator training. Integration of StyleGAN2 and encoder based mapping (coupling best of the two methods) offers a promising future direction to be investigated, to improve latent space mapping and thus enabling UAD in histopathological analysis.

\section{Conclusion}
\label{sec:conc}

In this work we addressed image synthesis as a pretext for GAN based anomaly detection pipeline in histopathological diagnosis and demonstrated, that histology imagery of high quality and variability can be synthesized, as well as reconstructed. We identified the importance of synthesizing large histology samples, not used in current GAN based anomaly detection pipelines, as well as the drawbacks and future research direction for more effective latent space mapping. The ability to generate realistically looking normal histology imagery of high resolution and size will enable the development of UAD pipeline, in order to apply it to cancer diagnosis, especially important for rare cancer types (e.g. paediatric), where annotated data is scarce, thus preventing the use of supervised approaches. Reducing the performance gap between supervised and unsupervised approaches and increasing the robustness of the UAD approaches will represent a significant contribution to wider adaption of automated visual analysis techniques, well beyond presented medical domain.

\section*{Acknowledgment}

This work was partially supported by the European Commission through the Horizon 2020 research and innovation program under grant 826121 (iPC).

\bibliographystyle{ieeetr}
\bibliography{literature}

\end{document}